\documentclass{article}
\usepackage{spconf,amsmath,epsfig}
\usepackage{caption}
\usepackage{subcaption}
\usepackage{enumitem} 
\usepackage[labelsep=period]{caption}
\renewcommand{\thetable}{\Roman{table}}
\usepackage[backend=biber]{biblatex}
\title{ A NEW PUBLIC ALSAT-2B DATASET FOR SINGLE-IMAGE SUPER-RESOLUTION}
\name{ Achraf Djerida, Khelifa Djerriri, Moussa Sofiane Karoui and Mohammed El Amin larabi     }
\address{ Agence Spatiale Algérienne, Centre des Techniques Spatiales, Arzew, Algérie \\  \{adjerida, kdjerriri, skaroui, mlarabi \}@cts.asal.dz}

\begin{document}
%
\maketitle
\begin{abstract}
 Currently, when reliable training datasets are available, deep learning methods dominate the proposed solutions for image super-resolution. However, for remote sensing benchmarks, it is very expensive to obtain high spatial resolution images. Most of the super-resolution methods use down-sampling techniques to simulate low and high spatial resolution pairs and construct the training samples. To solve this issue, the paper \footnote{Copyright 2021 IEEE. Published in the IEEE 2021 International Geoscience and Remote Sensing Symposium (IGARSS 2021), scheduled for July 11 - 16, 2021 in Brussels, Belgium. Personal use of this material is permitted. However, permission to reprint/republish this material for advertising or promotional purposes or for creating new collective works for resale or redistribution to servers or lists, or to reuse any copyrighted component of this work in other works, must be obtained from the IEEE. Contact: Manager, Copyrights and Permissions / IEEE Service Center / 445 Hoes Lane / P.O. Box 1331 / Piscataway, NJ 08855-1331, USA. Telephone: + Intl. 908-562-3966.} introduces a novel public remote sensing dataset (Alsat-2B) of low and high spatial resolution images (10m and 2.5m respectively) for the single-image super-resolution task. The high-resolution images are obtained through  pan-sharpening. Besides, the performance of some  super-resolution methods on the  dataset is assessed based on  common criteria. The obtained results reveal that the proposed scheme is promising and highlight the challenges in the dataset which shows the need for advanced methods to grasp the relationship between the low and high-resolution patches.
\end{abstract}
\section{Introduction}
\label{sec:intro}
 
  Image Super-Resolution (SR) approaches receives great attention due to its ability to generate a High-spatial-Resolution (HR) image from a Low-spatial-Resolution (LR) one without the need to other extra information  \cite{yang2014single}. This advantage is very crucial for many remote sensing applications that need to go beyond the sensor limits. There are mainly two types of image super-resolution: Single Image Super-Resolution  (SISR) and Multiple-Image Super-Resolution  (MISR). SISR methods reconstruct an HR image  from only a  single LR image, while LR multiple images of the same scene, through one or different sensors, are needed for MISR methods to provide an HR image that takes into account information involved in these multiple images. This work focuses on SISR techniques that are of great importance for many applications where the availability of LR multiple images of the same scene is not straightforward.

So far, there exist a large variety of methods aiming at solving  the SISR problem. In a rigorous work \cite{fernandez2017single}, it is shown that learning-based solutions tend to outperform other methods when high up-scaling factors are needed. Also, it is shown that when enough training examples are available, the learning-based methods are always preferred. On the other hand, if the training dataset is not large enough and does not reflect real challenges, reconstruction and hybrid methods are preferred. The lack of HR images is another issue for remote sensing applications, which is the main reason for the investigation of SR methods \cite{fernandez2017single}. A convolutional generator is proposed in \cite{haut2018new} to super-resolve remote sensing images based on an unsupervised way. The LR images are generated based on HR images through a procedure that includes blurring and a decimation process. Similarly, an unsupervised SR method is proposed in \cite{zhang2020unsupervised} without the need for HR images in the training phase. The generation of the training dataset is accomplished by the down-sampling of remote sensing images via  bicubic resampling. 

By examining most of the developed methods for remote sensing, it clearly appears that the creation of LR images is accomplished through down-sampling methods, and there are only very few  works that consider the generation of HR images with the help of other HR spectral bands. The main approach that takes the advantage of the Panchromatic (PAN) band is developed in  \cite{muller2020super} where pairs of LR multi-spectral (MS) and HR pan-sharpened images are created. In that work, the   method uses the PAN band of Pléiades  and aims at processing the images at full 12-bit information depth. A training set is constructed in \cite{galar2019super} for Sentinel-2 images to super-resolve 10m bands to higher resolutions based on other suitable sensors. The scheme is shown to be promising through the application of several deep convolutional networks.

This paper proposes an approach to super-resolve Alsat-2B images from 10m to 2.5m. It includes mainly two contributions:
\begin{enumerate}[label=(\roman*)]
\item The development of a novel dataset of LR/HR pairs based on Alsat-2B images. Rather than down-sampling methods, the HR images are generated based on the PAN band of Alsat-2B. The dataset incorporates several challenges that can help in developing SR methods to be more reliable.

\item The assessment of two lightweight deep learning methods on the Alsat-2B dataset with the Peak Signal to Noise Ratio (PSNR) and Structural SIMilarity (SSIM) criteria.

\end{enumerate}
  
   \begin{figure}[t]

  \centering
 \centerline{\epsfig{figure=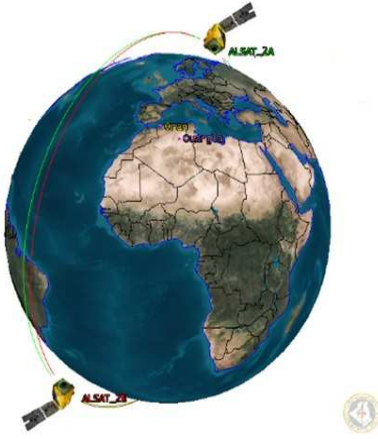,width=6cm,height=6cm}}

%
\caption{Orbits of Alsat-2B and Alsat-2A.}
\label{fig1}
\end{figure}

\section{Dataset Creation}
\label{sec:format}
\subsection{Alsat-2B mission}
Alsat-2B \footnote{https://asal.dz/} is an Earth observation satellite launched by the Algerian Space Agency in September 2016 to complement Alsat-2A, which was launched in 2010  (see Fig.~\ref{fig1}). It is equipped with the New AstroSat Optical Modular Instrument (NAOMI), a  payload that can supply images with a spatial resolution of 2.5 meters in the PAN mode and 10 meters in each of the four spectral bands (Red, Green, Blue and Near-InfraRed) in the MS mode. Its revisit time is around 3 days which permits it to be used for a variety of applications such as agriculture, disaster monitoring, and urban planning.

\subsection{Collection of satellite imagery}

To construct the dataset, 13 Alsat-2B images, which cover 13
different cities are used. The images are delivered under the
level 2A that means they are corrected for radiometric and
geometric errors, in addition, they are selected to contain different land-use classes with negligible clouds presence. Each
Alsat-2B product is composed of panchromatic and multi-
spectral images in the form of 12-bit unsigned integer.

\begin{table}[t]
\caption{Comparison between the Alsat-2B dataset and other common remote sensing datasets.}
\begin{center}
\begin{tabular}{|c|c|c|c|}
\hline
\textbf{Dataset}&\multicolumn{3}{|c|}{\textbf{Dataset properties}} \\
\cline{2-4} 
\textbf{} & \textbf{\textit{Size}}& \textbf{\textit{Resolution}}& \textbf{\textit{Patch size}} \\
\hline
UC  \cite{yang2010bag}& 2100&30cm & 256 \\
\hline
WHU\cite{dai2010satellite}& 1005& Up to 0.5m& 600 \\
\hline
NWPU\cite{cheng2017remote}& 31500& 30 to 0.2 m&256  \\
\hline
RSSCN7\cite{zou2015deep}& 2800&N/A & 400 \\
\hline
Alsat-2B&2759 pairs & around 2.5m & 256 \\
\hline
\end{tabular}
\label{tab1}
\end{center}
\end{table}

\subsection{Generation of high-resolution images}
To generate HR images, the advantage of the PAN spectral band, to pan-sharpen LR images, is used. Pan-sharpening is a well-known technique to enhance the spatial resolution of satellite imagery by fusing the information from the PAN and MS bands \cite{larabi2020multibranch}. In the first step, Regions Of Interests (ROIs), in each image, are determined and the bands are cut simultaneously using the QGIS 3.10 software. To get the pan-sharpened images, the Cubic Resampling algorithm, from the GDAL library, is adopted \footnote{https://docs.qgis.org/3.16/en/docs/}.


 
%

%

\begin{figure*}
     \centering
     \begin{subfigure}[b]{0.22\textwidth}
         \centering
         \includegraphics[width=\textwidth]{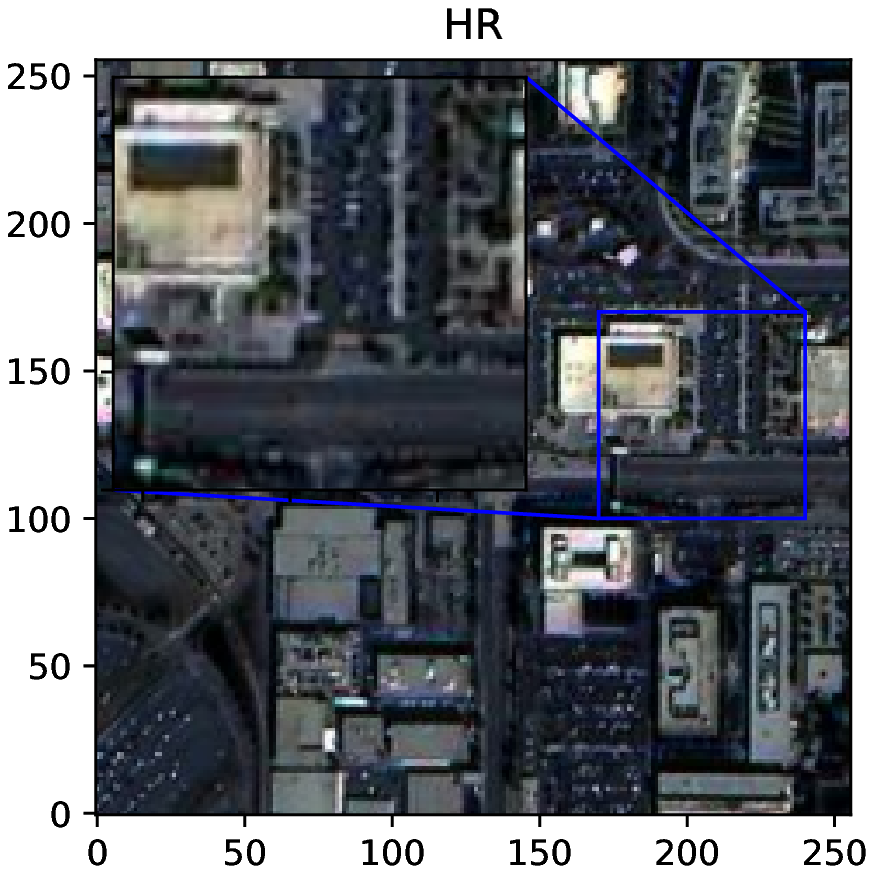}  
         \label{fig:y equals x}
     \end{subfigure}
    \hfill
     \begin{subfigure}[b]{0.22\textwidth}
         \centering
         \includegraphics[width=\textwidth]{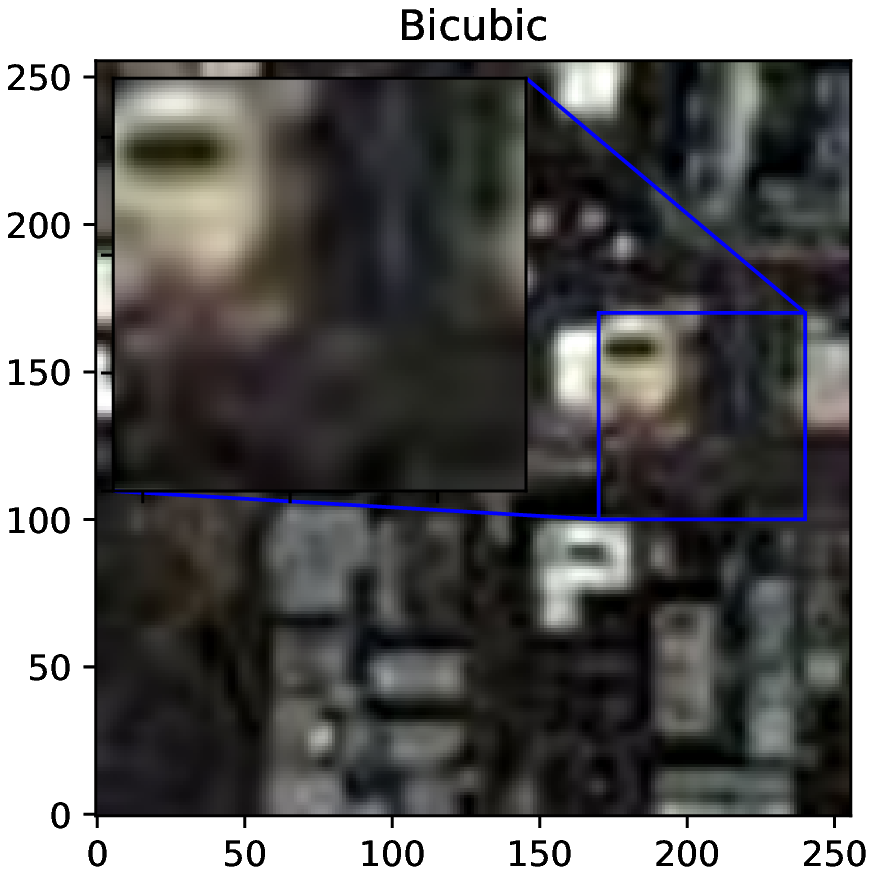}  
         \label{fig:three sin x}
     \end{subfigure}
     \hfill
     \begin{subfigure}[b]{0.22\textwidth}
         \centering
         \includegraphics[width=\textwidth]{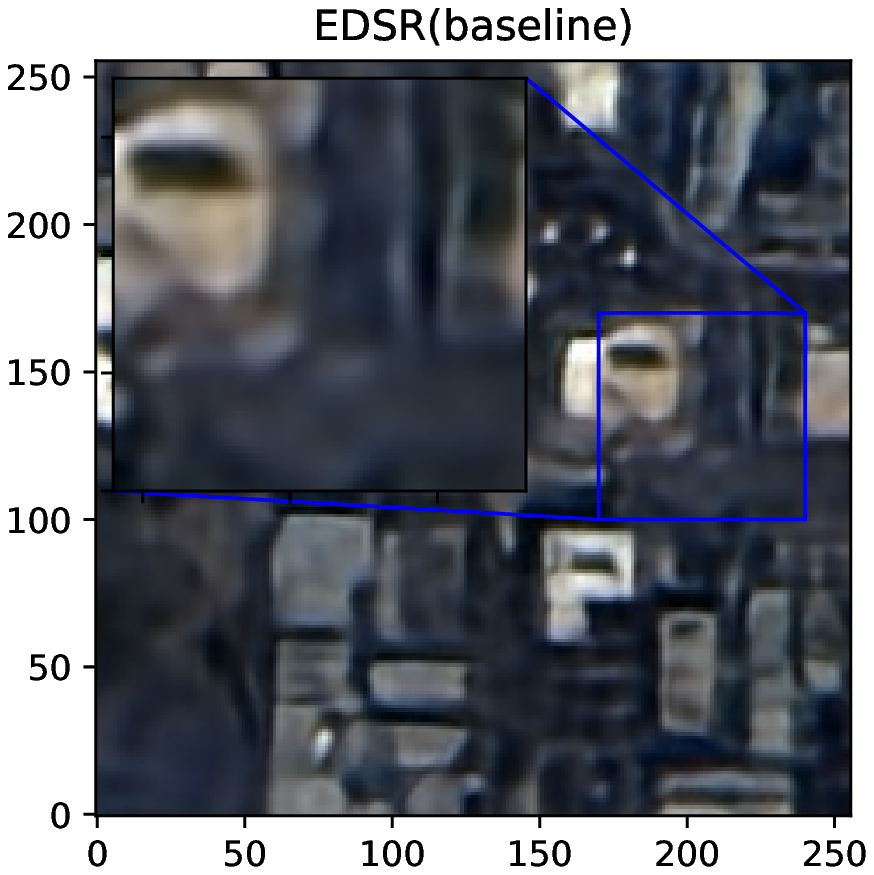}  
         \label{fig:five over x}
     \end{subfigure}
     \hfill
     \begin{subfigure}[b]{0.22\textwidth}
         \centering
         \includegraphics[width=\textwidth]{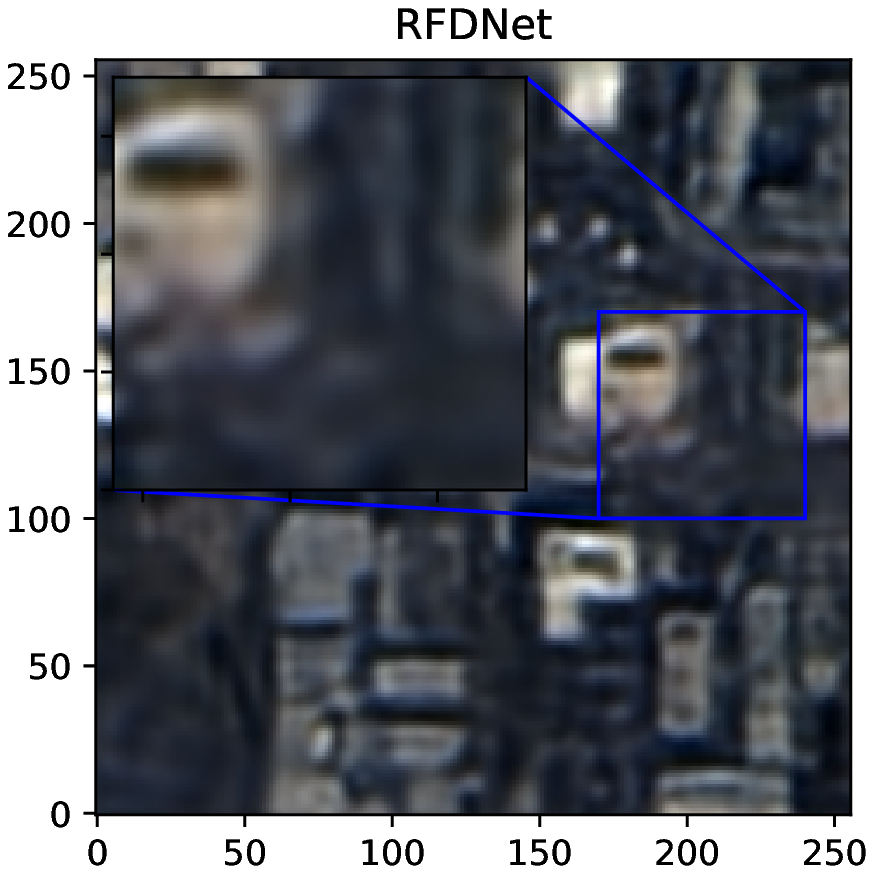}  
         \label{fig:five over x}
     \end{subfigure}
     
        \caption{A qualitative comparison of some SR methods on a sample patch from Alsat-2B.}
        \label{fig2}
\end{figure*}

\subsection{Generation of training and test patches}

As deep learning methods are trained on patches, both the LR/HR images need to be transformed into patches of equal size. The patch size is fixed as 64$\times$64-pixel for the LR images and hence 256$\times$256-pixel for the HR ones, which reflects scaling factor of 4. At such scaling factor and spatial resolution, it is clear that grasping the relationship between the low and high spatial resolution patches is becoming a challenging task. To avoid a biased dataset, the training and test patches are not overlapped that means they do not cover the same geographical area.

Furthermore, the dataset size is kept manageable to verify the immunity of the patches toward other effects such as pan-sharpening limitations and cloud cover. The patches cover mainly three classes: urban, agriculture, and special structures, which highlight distinguished objects such as stadiums and bridges. In total, the dataset is composed of 2759 pairs of LR/HR patches where the training set size is 2182 and the test set size is 577. To furthermore evaluate the effect of the patch content on the performance of the super-resolution methods, the test set is divided into three sub-classes as explained before: urban (239), agriculture (56), and special (282). The next table (Table \ref{tab1}) shows a comparison between the developed and common datasets that are used in the evaluation of SR methods. It is well-noted that the mentioned datasets are dedicated originally to scene classification. However, they are used often as HR images and down-sampled to create their corresponding LR samples. Hence, the developed dataset overcomes this limitation and helps in a more reliable evaluation.

To make the images suitable for most deep learning methods, they are converted to RGB with 8-bit unsigned integer format and it is made available publically\footnote{https://github.com/achrafdjerida/Alsat-2B}.

\section{SINGLE-IMAGE SUPER-RESOLUTION OF ALSAT-2B IMAGES}

 So far in the literature, there exist a large number of proposed deep learning methods varying from low to high complex architectures that need powerful GPUs and excessive time during the training process. Although complex models and excessive training  are essential for challenging problems, the assessment of lightweight models is the first step to be done to reveal the level of difficulty in the dataset. Therefore, two state-of-the-art lightweight deep learning architectures Residual Feature Distillation Network (RFDNet)  \cite{liu2020residual} and and the baseline of the Enhanced Deep Super-Resolution network (EDSR)  \cite{lim2017enhanced} are considered to super-resolve the used Alsat-2B images. RFDNet employs the Feature Distillation Connection (FDC) to leverage the heavy computations of Convolutional Neural Network (CNN) models. The design of the method aims at benefiting from residual learning while keeping the model lightweight enough. EDSR is a high-level architecture and a winner of the NTIRE 2017 super-resolution challenge. Its residual block design does not include batch normalization and the final ReLU activation. To leverage the computations, the baseline version of EDSR is adopted since it has a lighter structure due to the removal of residual scaling layers.

\section{EXPERIMENTS}
\subsection{Implementation details}
The public Keras/Tensoflow implementations of RFDNet\footnote{ https://paperswithcode.com/paper/residual-feature-distillation-network-for} and EDSR\footnote{https://github.com/krasserm/super-resolution} are used. RFDNet is trained with ADAM optimizer (learning rate $5.10^{-5}$) and the Mean Square Error (MSE) as the loss function.  EDSR is trained with ADAM optimizer (learning rate $10^{-4}$) to minimize the Mean Absolute Error (MAE) with 16 residual blocks. Both methods are trained for 500 epochs, batch size 8 and up-scale fator 4. All the experiments are performed in the Google Colab Cloud GPU.

\subsection{Qualitative results }
 
 The next figure (Fig.~\ref{fig2}) shows a qualitative comparison on a sample scene from the Alsat-2B dataset. Clearly, both RFDNet and EDSR shows better spatial details than LR and the Bicubic interpolation. The blurring effects which are highly present in the LR images are reduced and the edges of different objects become more sharpened. However, it is evident that still much room for improvements so that the SR results are closer to HR ones.

\begin{table*}[t]
\caption{Comparison of tested SR methods on the Alsat-2B dataset based on the PSNR / SSIM criteria.}
\begin{center}
\begin{tabular}{|c|c|c|c|c|}
\hline
\textbf{Tested SR Methods}&\multicolumn{4}{|c|}{\textbf{Test dataset}} \\
\cline{2-5} 
\textbf{} & \textbf{\textit{Special}}& \textbf{\textit{Agriculture}}& \textbf{\textit{Urban}} &\textbf{Mean}\\ 
\hline
Lanczos3 &14.86/0.2822 &14.59/0.3376&13.27/0.2140&14.24/0.2779  \\
\hline
Bicubic &14.98/0.2829 &14.68/0.3418&13.41/0.2128&14.35/0.2791  \\
\hline
EDSR(baseline) &16.35/\textbf{0.3569} &17.24/\textbf{0.4381}&14.88/\textbf{0.2904}&16.16/\textbf{0.3618}  \\
\hline
RFDNet &\textbf{16.63}/0.3528 &\textbf{17.32}/0.4283&\textbf{15.17}/0.2835 & \textbf{16.37}/0.3548\\
\hline
\end{tabular}
\label{tab2}
\end{center}
\end{table*}

\subsection{Quantitative  results }

To assess the performance of SR methods, the PSNR and SSIM are computed and reported in the following table (Table ~\ref{tab2}). The highest values of these criteria correspond to the best SR results. Clearly, the learning-based methods achieve the best results based on both criteria. However, the overall results are still low that reflect some challenges in the dataset.

\subsection{Discussion }

Although the learning-based methods show better performance compared to other methods and especially  improved spatial details compared to LR images, the obtained statistics demonstrate that the generated SR images are still far from the reference HR images. This remark demonstrates the challenges in the Alsat-2B dataset due to many factors especially the high scaling factor compared to the spatial resolution of the LR images. The performance can be boosted by the use of advanced techniques such as data augmentation and complex models.

\section{Conclusion}

In this paper, a novel Alsat-2B dataset, for the single-image super-resolution task, is introduced. The dataset is developed to avoid the use of down-sampling methods and it provides an opportunity to evaluate SR methods on  high spatial resolution images created via a pan-sharpening process. Two  deep-learning methods are evaluated based on common criteria. Although results show improved performance compared to interpolation methods, it is clear that more advanced techniques are needed to grasp the relationship between the pairs. As future work, data augmentation and complex models will be adopted to super-resolve the Alsat-2B images.

\printbibliography

\end{document}